\documentclass[times, 10pt, twocolumn]{article} 
\usepackage{times}
\usepackage{epsfig}

\newcommand{\ie}{{\it i.e.,$\ $}}
\newcommand{\eg}{{\it e.g.,$\ $}}

\begin{document}

\title{Proactive Service Migration for Long-Running Byzantine Fault Tolerant Systems\thanks{This research has been
       supported in part by Department of Energy 
Contract DE-FC26-06NT42853, and by Cleveland State University 
through a Faculty Research Development award.}}
       \author{Wenbing Zhao \\
       Department of Electrical and Computer Engineering \\ Cleveland
       State University, 2121 Euclid Ave., Cleveland, OH 44115\\
       wenbing@ieee.org} \date{} \maketitle
\thispagestyle{empty}

\begin{abstract}
In this paper, we describe a novel proactive recovery scheme based on service 
migration for long-running Byzantine fault tolerant systems. Proactive 
recovery is an essential method for ensuring long term reliability of 
fault tolerant systems that are under continuous threats from malicious 
adversaries. The primary benefit of our proactive recovery scheme is a 
reduced vulnerability window. This is achieved by removing the time-consuming 
reboot step from the critical path of proactive recovery. Our migration-based
proactive recovery is coordinated among the replicas, therefore, it can 
automatically adjust to different system loads and avoid the problem of
excessive concurrent proactive recoveries that may occur in previous work
with fixed watchdog timeouts. Moreover, the fast proactive recovery also 
significantly improves the system availability in the presence of faults.
\end{abstract}



\noindent
{\bf Keywords:} Proactive Recovery, Byzantine Fault Tolerance, Service Migration, Replication, Byzantine Agreement

\section{Introduction}

We have seen increasing reliance on services provided over the Internet. 
These services are expected to be continuously available over extended 
period of time (typically 24x7 and all year long). Unfortunately,
the vulnerabilities due to insufficient design and poor implementation 
are often exploited by adversaries to cause a variety of damages, e.g., 
crashing of the applications, leaking of confidential information, 
modifying or deleting of critical data, or injecting of erroneous 
information into the application data. These malicious faults are often 
modeled as Byzantine faults~\cite{lamport:byz}, 
and they are detrimental to any online
service providers. Such threats can be coped with using Byzantine
fault tolerance (BFT) techniques, as demonstrated by many research results
\cite{bft-osdi99, bft-osdi2000, bft-acm, base,hq,thema,oceanstore,alvisi-bft}. 
The Byzantine fault tolerance algorithms assume that only
a small portion of the replicas can be faulty. When the number of faulty
replicas exceeds a threshold, BFT may fail. Consequently, Castro and
Liskov \cite{bft-osdi2000} proposed a proactive recovery scheme that
periodically reboots replicas and refreshes their state, even before
it is known that they have failed. As long as the number of compromised
replicas does not exceed the threshold within a time window that all replicas
can be proactively recovered (such window is referred to as window of 
vulnerability~\cite{bft-acm}, or vulnerability window), 
the integrity of the BFT algorithm holds and the services being protected 
remain highly reliable over the long term.

However, the reboot-based proactive recovery scheme has a number of
issues. First, it assumes that a simple reboot (\ie power cycle the computing
node) can successfully repair a compromised node, which might not be the case, 
as pointed out in \cite{bftlls}. Second, even if a compromised
node can be repaired by a reboot, it is often a prolonged
process (typically over 30$s$ for modern operating systems). During 
the rebooting step, the BFT services might not be available
to its clients (\eg if the rebooting node happens to be a nonfaulty replica
needed for the replicas to reach a Byzantine agreement). Third, there lacks 
coordination among replicas to ensure that no more than a small portion of the 
replicas (ideally no more than $f$ replicas in a system of $3f+1$ replicas to 
tolerate up to $f$ faults) are undergoing proactive recovery at any 
given time, otherwise, the services may be unavailable for extended 
period of time. The static watchdog 
timeout used in \cite{bft-acm} also contributes to the problem because it
cannot automatically adapt to various system loads. The staggered 
proactive recovery scheme in \cite{bft-acm} is not sufficient to prevent
this problem from happening.

In this paper, we present a novel proactive recovery scheme based
on service migration, which addresses all these issues. 
Our proactive recovery scheme requires the availability
of a pool of standby computing nodes in addition to the active nodes 
where the replicas are deployed. The basic idea is outlined below.
Periodically, the replicas initiate a proactive recovery by selecting a 
set of active replicas, and a set of target standby nodes for a
service migration. At the end of the service migration, the source
active nodes will be put under a series of preventive sanitizing and repairing
steps (such as rebooting and swapping in a clean hard drive with the 
original system binaries) before they are assigned to the pool of 
standby nodes, and the target nodes are promoted to the group of active 
nodes. The unique feature of this design is that the sanitizing and 
repairing step is carried out {\em off the critical path of proactive 
recovery} and consequently, it has minimum negative impact on the 
availability of the services being protected.

This paper makes the following research contributions: 
\begin{itemize}
\item We propose a novel migration-based proactive recovery scheme for 
long-running Byzantine fault tolerant systems. The scheme significantly reduces
the recovery time, and hence, the vulnerability window, by moving the
time-consuming replica sanitizing and repairing step off the critical path.

\item Our proactive recovery scheme ensures a coordinated periodical
recovery, which prevents harmful excessive concurrent proactive recoveries.

\item We present a comparison study on the performance of the reboot-based 
and our migration-based proactive recovery schemes in the presence
of faults, both by analysis and by experiments.

\end{itemize}

\section{System Model}

We assume a partially asynchronous distributed system in that all
message exchanges and processing related to proactive recovery can
be completed within a bounded time. This bound can be initially set
by a system administrator and can be dynamically adjusted by the
recovery mechanisms. However, the safety property of the Byzantine 
agreement on all proactive recovery related decisions (such as
the selection of source nodes and destination nodes for service
migration) is maintained without any system synchrony requirement.

We assume the availability of a pool of nodes to serve as the standby
nodes for service migration, in addition to the $3f+1$ active nodes required 
to tolerate up to $f$ Byzantine faulty replicas. The pool size is large enough
to repair damaged nodes while enabling frequent service
migration for proactive recovery. Furthermore, both active nodes
and standby nodes can be subject to malicious attacks (in addition
to other non-malicious faults such as hardware failures). However,
we assume that the rate of successful attacks on the standby nodes
is much smaller than that on active nodes, \ie the tolerated
successful attack rate on active nodes is determined by the 
vulnerability window the system can achieve, and the tolerated 
successful attack rate on standby nodes is determined by the repair
time. The allowed repair time can be much larger than the
achievable vulnerability window given a sufficiently large pool of standby
nodes. If the above assumptions are violated,
there is no hope to achieve long-term Byzantine fault tolerance.

We assume the existence of a trusted configuration manager, as described in
\cite{rosebud,bftlls}, to manage the pool of standby nodes, and to assist 
service migration. Example tasks include
frequently probing and monitoring the health of each standby node, and 
repairing any faulty node detected.
We will not discuss the mechanisms used by the manager to carry out
such tasks, they are out of the scope of this paper. 

Other assumptions regarding the system is similar to those in \cite{bft-acm}
and they are summarized here. All communicating entities (clients,
replicas and standby nodes) use a secure hash function such as SHA1 to compute
the digest of a message and use the message authentication codes (MACs)
to authenticate messages exchanged, except for key exchange messages, which
are protected by digital signatures. For point to point message exchanges,
a single MAC is included in each message, while multicast messages are
protected by an authenticator \cite{authenticator}. Each entity has a pair 
of private and public keys. The active and standby nodes each is equipped with
a secure coprocessor and sufficiently large read-only memory. In these
nodes, the private key is stored in the coprocessor and all digital
signing and verification is carried out by the coprocessor without revealing
the private key. The read-only memory is used to store the execution code
for the server application and the BFT framework. We do not require the 
presence of a hardware watchdog timer because of the coordination of 
migration and the existence of a trusted configuration manager.

Finally, we assume that an adversary is computational bound so that it cannot 
break the above authentication scheme. 

\section{Proactive Service Migration Mechanisms}

The proactive service migration mechanisms collectively ensure the
following objectives:
\begin{enumerate}
\item To ensure that correct active replicas have a consistent
membership view of the available standby nodes.

\item To determine when to migrate and how to initiate a migration.

\item To determine the set of source and target nodes for migration.

\item To transfer a correct copy of the system state to the new replicas.

\item To notify the clients the new membership after each proactive recovery.
\end{enumerate}
The first objective is clearly needed because otherwise the replicas
cannot possibly agree on the set of target nodes for migration. The
second and third objectives are critical to ensure a coordinated periodic 
proactive recovery. The fourth objective is obviously necessary for the new 
replicas to start from a consistent state. The fifth objective is essential to
ensure that the clients know the correct membership of the server replicas
so that they do not accept messages from possibly faulty replicas that
have been migrated out of active executing duty, and they can send
requests to the new replicas.

\subsection{Standby Nodes Registration}
\begin{figure}[t]
\begin{center} 
\leavevmode
\epsfxsize=3.0in
\epsfbox{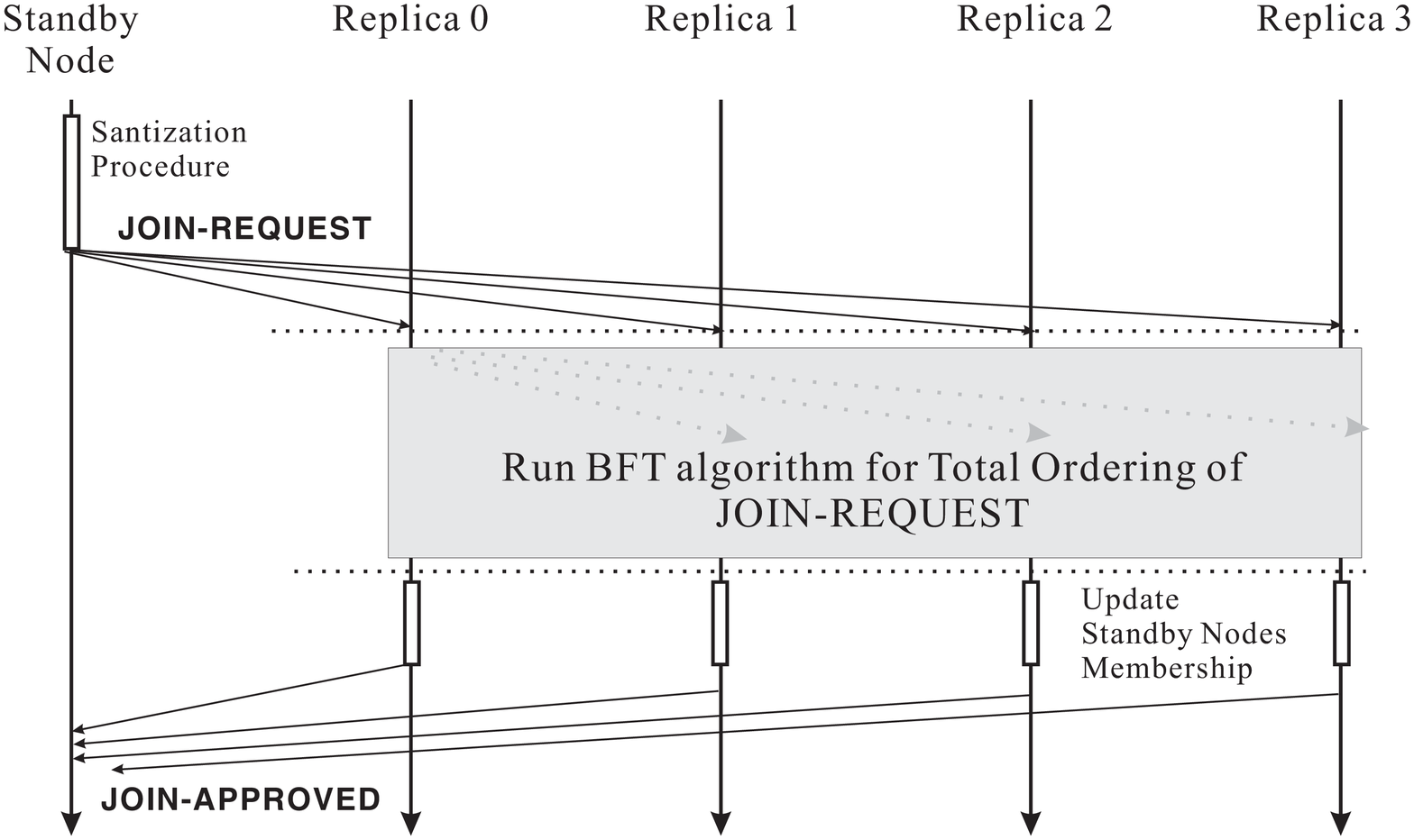}
\caption{The protocol used for a standby node to register with active
replicas.}
\label{joinfig}
\end{center}
\end{figure}

Each standby node is controlled by the trusted configuration manager
and is undergoing constant probing and sanitization procedures such
as reboot. If the configuration manager suspects the node to be faulty
and cannot repair it automatically, a system administrator might be
called in to manually fix the problem. Each time a standby node completes
a sanitization procedure, it notifies the active replicas with a 
{\sc join-request} message in the form of 
$<${\sc join-request}$, l, i_s$$>$$_{\sigma_{i_s}}$, where $l$ is the counter 
value maintained by the secure coprocessor of the standby node, $i_s$ 
is the identifier of the standby node, and $\sigma_{i_s}$ is the 
authenticator. The registration protocol is illustrated 
in Figure~\ref{joinfig}.

An active replica accepts the {\sc join-request} if it has not
accepted one from the same standby node with the same or greater $l$.
The {\sc join-request} message, once accepted by the primary, is ordered the 
same way as a regular message with a sequence number $n_r$, except that
the primary also assigns a timestamp as the join time of the standby node
and piggybacks it with the ordering messages.
The total ordering of the {\sc join-request} is important so that all
active nodes have the same membership view of the standby nodes.
The significance of the join time will be elaborated later in this section.

When a replica executes the {\sc join-request} message, it sends a 
{\sc join-approved} message in the form of 
$<${\sc join-approved}$, l, n_r$$>$$_{\sigma_i}$ to the
requesting standby node. The requesting standby node must collect
$2f+1$ consistent {\sc join-approved} messages with the same $l$ and $n_r$
from different active replicas. The standby node then initiates a key
exchange with all active replicas for future communication.

A standby node might go through multiple rounds of proactive sanitization
before it is selected to run an active replica. The node sends a 
new {\sc join-request} reconfirming its membership after each round of
sanitization. The active replicas subsequently updates the join time
of the standby node.

It is also possible that the configuration manager deems a registered
standby node as faulty and it requires a lengthy repair, in which case,
the configuration manager deregisters the faulty node from active
replicas by sending a {\sc leave-request}. The {\sc leave-request} 
is handled by the active replicas in a similar way as that for
{\sc join-request}. In the unlikely case that the faulty standby node has 
been selected as the new active node, the mechanisms react in the following
ways: (1) if the migration is still ongoing when the {\sc leave-request} 
arrives, it is aborted and restarted with a different set of target standby 
nodes, and (2) if the migration has been completed, an on-demand
service migration will be initiated to swap out the faulty node. 
The on-demand service migration mechanism is rather similar to the proactive 
migration mechanism, as will be discussed in Section~\ref{ondemandsec}.

\subsection{Proactive Service Migration}
\paragraph{When and How to Initiate a Proactive Service Migration?}
The proactive service migration is triggered by the software-based
migration timer maintained by each replica. The timer is reset and
restarted at the end of each round of migration. (An on-demand service 
migration may also be carried out upon the notification from the
configuration manager, as mentioned in the previous subsection.)

How to properly initiate a proactive service migration, however, is
tricky. We cannot depend on the primary to initiate a proactive
recovery because it might be faulty. Therefore, the migration initiation
must involve all replicas.

On expiration of the migration timer, a replica chooses a set of $f$ active
replicas, and a set of $f$ standby nodes, and multicasts an 
{\sc init-migration} request to all other replicas in the form 
$<${\sc init-migration}$, v, l, S, D, i$$>$$_{\sigma_i}$, where $v$ is the 
current view, $l$ is the migration number (determined by the number of 
successful migration rounds recorded by replica $i$), $S$ is the set of 
identifiers for the $f$ active replicas to be migrated, $D$ is the set of 
identifiers for the $f$ standby nodes as the targets of the migration,
$i$ is the identifier for the sending replica, and $\sigma_i$ is the
authenticator for the message.

On receiving an {\sc init-migration} message, a replica $j$ accepts the message
and stores the message in its data structure provided that the message
carries a valid authenticator, it has not accepted an {\sc init-migration}
message from the same replica $i$ in view $v$ with the same or higher
migration number, and the replicas in $S$ and $D$ are consistent with
the sets determined by itself according to the selection algorithm
(to be introduced next).

\begin{figure}[t]
\begin{center} 
\leavevmode
\epsfxsize=3.0in
\epsfbox{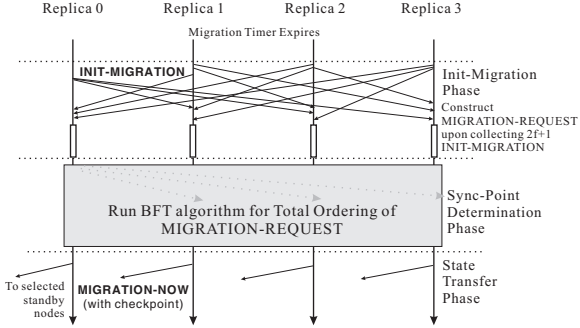}
\caption{The proactive service migration protocol.}
\label{migrationfig}
\end{center}
\end{figure}

Each replica waits until it has collected $2f$$+$$1$ {\sc init-migration} 
messages from different replicas (including its own {\sc init-migration} 
message) before it constructs a {\sc migration-request} message. 
The {\sc migration-request} message has
the form $<${\sc migration-request}$, v, l, S, D$$>$$_{\sigma_p}$. The
primary, if it is correct, should place the {\sc migration-request}
message at the head of the request queue and order it immediately.
The primary orders the {\sc migration-request} in the same way as that for a 
normal request coming from a client, except that (1) it does not batch 
the {\sc migration-request} message with normal requests,
and (2) it piggybacks the {\sc migration-request} and the
$2f$$+$$1$ {\sc init-migration} messages (as proof of validity of the
migration request) with the {\sc pre-prepare} message.
The reason for ordering the {\sc migration-request} is to ensure
a consistent synchronization point for migration at all replicas.
An illustration of the migration initiation protocol is shown as part of 
Figure~\ref{migrationfig}.

Each replica starts a view change timer when the {\sc migration-request}
message is constructed (just like when it receives a normal request) so 
that a view change will be initiated if the primary is faulty and does not 
order the {\sc migration-request} message. The new primary, if it is
not faulty, should continue this round of proactive migration.

In this work, we choose not to initiate a view change when the primary
is migrated if the state is smaller than a tunable parameter (100KB
is used in our experiment). For larger state (\ie when the cost of state
transfer is more than that of the view change), the primary multicasts a 
{\sc view-change} message before it is migrated, similar to \cite{bft-acm}.

\paragraph{Migration Set Selection.}
The selection of the set of active replicas to be migrated is relatively
straightforward.
It takes four rounds of migration (each round for $f$ replicas) to 
proactively recover all active replicas at least once. The replicas
are recovered according to the reverse order of their identifiers, similar to
that used in \cite{bft-acm}. For example,
for the very first round of migration, replicas with identifiers
of $3f,3f-1,...,2f+1$ will be migrated, and this will be followed by replicas
with identifiers of $2f,2f-1,...,f+1$ in the second round, replicas
with identifiers of $f,f-1,...,1$ in the third round, and finally
replicas with identifiers of $0,3f,...2f+2$ in the fourth round. 
(The example assumed $f>2$. It is straightforward to derive the selections
for the cases when $f=1,2$.)
The selection is deterministic and can be easily computed
based on the migration number. Note that the migration number constitutes
part of middleware state and will be transfered to all recovering
replicas. The selection is independent of the view the
replicas are in.

The selection of the set of standby nodes as the target of migration
is based on the elapsed time since the standby nodes were last sanitized.
That is why each replica keeps track of the join time of each standby
nodes. For each round of migration, the $f$ standby nodes
with the least elapsed time will be chosen. This is out of consideration
that the probability of these nodes to be compromised at the time of
migration is the least (assuming brute-force attacks by adversaries).

\paragraph{Migration Synchronization Point Determination.}
It is important to ensure all (correct) replicas to use the same 
synchronization point when performing the service migration. This is
achieved by ordering the {\sc migration-request} message. 
The primary starts to order the message by sending a {\sc pre-prepare} 
message for the {\sc migration-request} to all backups, as described 
previously. 

A backup verifies the piggybacked {\sc migration-request} in a similar fashion
as that for the {\sc init-migration} message, except now the replica must check
that it has received all the $2f$$+$$1$ init-migration messages that the
primary used to construct the {\sc migration-request}, and the sets in
 $S$ and $D$ match those in the {\sc init-migration} messages. The backup
requests the primary to retransmit any missing {\sc init-migration} messages.
The backup accepts the {\sc pre-prepare} message for the 
{\sc migration-request} provided that the {\sc migration-request} is correct 
and it has not accepted another {\sc pre-prepare} message for the same 
sequence number in view $v$. From now on, the replicas executes according to 
the three-phase BFT algorithm~\cite{bft-acm} as usual 
until they commit the {\sc migration-request}.

\paragraph{State Transfer.}
When it is ready to execute the {\sc migration-request}, a replica $i$ takes 
a checkpoint of its state (both the application and the BFT middleware state), 
and multicasts a {\sc migrate-now} message to the $f$ standby nodes selected. 
The {\sc migrate-now} message has the form 
$<${\sc migrate-now}$, v, n, C, P, i$$>$$_{\sigma_i}$, where $n$ is 
the sequence number assigned to the {\sc migration-request},
$C$ is the digest of the checkpoint, and $P$ contains $f$ tuples. 
Each tuple contains the identifiers of a source-node and target-node pair 
$<$$s, d$$>$. The standby node $d$, once completes the proactive recovery 
procedure, assumes the identifier $s$ of the active node it replaces.
A replica sends the actual checkpoint (together with all queued request 
messages, if it is the primary) to the target nodes in separate messages.

If a replica belongs to the $f$ nodes to be migrated, it performs the
following additional actions:
(1) it stops accepting new request messages, and 
(2) it reports to the trusted configuration manager as a candidate standby 
node.
This replica is then handed over to the control of the configuration
manager for sanitization.

Before a standby node can be promoted to run an active replica,
it must collect $2f$$+$$1$ consistent {\sc migrate-now} messages with the
same sequence number and the digest of the checkpoint from different active 
replicas. Once a standby node obtains a stable checkpoint, it applies the
checkpoint to its state and starts to accept clients' requests and
participate the BFT algorithm as an active replica.

\subsection{New Membership Notification}
One can envisage that a fault node might want to continue sending
messages to the active replicas and the clients, even if it has been
migrated, before it is sanitized by the configuration manager. It is
important to inform the clients the new membership so that they
can ignore such messages sent by the faulty replica. The membership
information is also important for the clients to accept messages
send by new active replicas, and to send messages to these replicas. 
This is guaranteed by the new membership notification mechanism.

The new membership notification is performed in a lazy manner to
improve the performance unless a new active replica assumes the primary
role, in which case, the notification is sent immediately to all known
clients (so that the clients can send their requests to the new primary). 
Furthermore, the notification is sent only by the existing active 
replicas (\ie not the new active replicas because the clients do not know 
them yet).
Normally, the notification is sent to a client only after the client has
sent a request that is ordered after the {\sc migration-request} message,
\ie the sequence number assigned to the client's request is bigger than
that of the {\sc migration-request}.

The notification message has the form 
$<${\sc new-membership}$, v, n, P, i$$>$$_{\sigma_i}$ (basically the same
as the {\sc migration-now} message without the checkpoint), where
$v$ is the view in which the migration occurred, and $n$ is the
sequence number assigned to the {\sc migration-request}, and $P$
contains the tuples of the identifiers for the replicas in the previous and 
the new membership. Note all active replicas should have the information.

When a client collects $f+1$ consistent {\sc new-membership} messages
from different replicas, it updates its state accordingly and starts
to accept replies from, and to send requests to, the new replicas.

\subsection{On-Demand Migration} 
\label{ondemandsec}
One demand migration can happen when the configuration manager detects a 
node to be faulty after it has been promoted to run an active replica. 
It can also happen when replicas have collected solid evidence that one or
more replicas are faulty,
such as a lying primary. The on-demand migration mechanism is rather
similar to that for proactive recovery, with only two differences:
(1) The migration is initiated on-demand, rather than by a migration timeout.
However, replicas still must exchange the {\sc init-migration} messages
before the migration can take place; (2) The selection procedure for the
source node is omitted because the nodes to be swapped out are already decided,
and the same number of target nodes are selected accordingly.

\section{Benefits of Proactive Service Migration}
\label{benefitsec}
\subsection{Reduced Vulnerability Window}
The primary benefit of using the migration-based proactive recovery 
is a reduced vulnerability window. The term vulnerability window (or window of 
vulnerability) $T_v = 2T_k + T_r$ is introduced in \cite{bft-acm}. 
Here $T_r$ is the time elapsed between when a replica becomes faulty and 
when it fully recovers from the fault, and $T_k$ is the key refreshment 
period. As long as no more than $f$ replicas
become faulty during the window of $T_v$, the invariants for Byzantine
fault tolerance will be preserved.

\begin{figure}[t]
\begin{center} 
\leavevmode
\epsfxsize=3.0in
\epsfbox{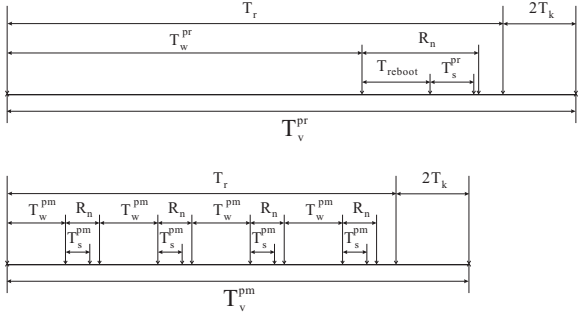}
\caption{Various time intervals used in modelling the vulnerability
windows for the two proactive recovery schemes.}
\label{windowfig}
\end{center}
\end{figure}

In the reboot-based proactive recovery scheme, the vulnerability window
$T_v^{pr}$ is characterized to be $2T_k + T_w^{pr} + R_n$, as shown in the 
upper half of Figure~\ref{windowfig}, where $T_w^{pr}$ is the watchdog timeout,
$R_n$ is the recovery time for a nonfaulty replica under normal load
conditions. The dominating factors for recovery time include $T_{reboot}$, the
reboot time, and $T_s^{pr}$, the time it takes to save, restore and verify
the replica state. The watchdog timeout $T_w^{pr}$ is set roughly to
$4R_n$ to enable a staggered proactive recovery of $f$ replicas at a time.

The composition of the vulnerability window for the migration-based 
proactive recovery is shown in the lower half of Figure~\ref{windowfig}.
The time intervals specific to migration-based proactive recovery is labeled
by the $pm$ superscript. Because the migration is coordinated in this
recovery scheme, no watchdog timer is used and the term $T_w^{pm}$ is
now interpreted as the migration timer, \ie the time elapsed between
two consecutive rounds of migrations of $f$ replicas each. This is
very different from the watchdog timeout $T_w^{pr}$, which is statically
configured prior to the start of each replica. Because the recovery time
in the migration-based proactive recovery is much shorter than that
in the reboot-based recovery, and the migration is coordinated,
it takes much shorter time to fully recovery all active replicas once.
Hence, $T_r$ can be much shorter for the migration-based recovery,
which leads to a smaller vulnerability window.

\subsection{Increased Availability in the Presence of Faults}
Under fault-free condition, neither the reboot-based nor the migration-based
recovery scheme has much negative impact to the runtime performance unless
the state is very large, as shown in the experimental data in~\cite{bft-acm} 
and in Section~\ref{perfsec} of this paper. However, in the presence of faulty 
nodes, the system availability can be reduced significantly in the 
reboot-based proactive recovery scheme, while the reduction in availability 
remains small in our migration-based recovery scheme.

The see the benefit of the migration-based proactive recovery regarding
the system availability in the presence of faults, we consider a 
specific case when the number of faulty nodes is $f$ and $f=1$. 
While developing a thorough analytical model is certainly
desirable, it is out of the scope of this paper.

We assume that there are $f=1$ faulty replica at the
beginning of the set of four rounds of migration to eradicate it.
(Recall that we assume that at most $f$ replicas can be compromised
in one vulnerability window, which constitutes four rounds proactive
recovery of $f$ replicas at a time and $2T_k$,
therefore, it is not possible to end up with more than $f$ faulty replicas
with this assumption.) We further assume that the proactive recovery
rounds after the removal of the faulty replica has no negative
impact on the system availability, and so does the case when the
faulty replica is recovered in the same round of recovery.
We also ignore the differences between
the recovery time of a normal replica and that of a faulty one.

Since $f=1$, the faulty node must be recovered in one of the four rounds
of recovery. Assuming that the faulty node is chosen randomly, it is
recovered in even probability in either of the four rounds, \ie
$P_i=0.25$, where $i=0,1,2,3$.
If the faulty replica is recovered in the first round of recovery,
there is no reduction of system availability $q_0$ (\ie $q_0=1$).
If the faulty replica is recovered in round $i$, where $i=1,2,3$,
the system will not be available while a replica is recovering during
each round because
there will be insufficient number of correct replicas until the recovery
is completed, and hence, the system availability $q_i$ in this case will
be determined as 
\begin{equation}
q_i=P_i\frac{T_v-iR_n}{T_v}
\end{equation}
Therefore, the total system availability is 
\begin{equation}
q=\sum_{i=0}^{3}q_i=0.25\sum_{i=0}^{3}\frac{T_v-iR_n}{T_v}
\end{equation}

\begin{figure}[t]
\begin{center} 
\leavevmode
\epsfxsize=3.0in
\epsfbox{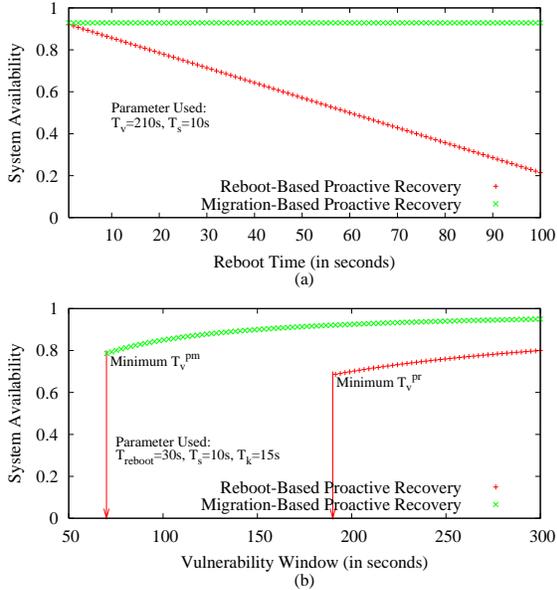}
\caption{An analytical comparison of the system availability between the 
reboot-based proactive recovery and the migration-based recovery. 
(a) With a fixed vulnerability window of 210$s$ and various reboot time. 
(b) With fixed recovery time and various vulnerability window settings.}
\label{avaifig}
\end{center}
\end{figure}

For the reboot-based recovery, $R_n\approx T_{reboot}+T_{s}^{pr}$, and
for the migration-based recovery, $R_n\approx T_{s}^{pm}$. It is not
unreasonable to assume $T_s^{pr}\approx T_s^{pm}$ because the network
bandwidth is similar to the disk IO bandwidth in modern general-purpose
systems. As shown in Figure~\ref{avaifig}(a), the migration-based recovery can 
achieve much better system availability if the reboot time $T_{reboot}$ is 
large, which is generally the case. Furthermore, as indicated in 
Figure~\ref{avaifig}(b), for the range of vulnerability window considered,
the system availability is consistently higher for the migration-based 
proactive recovery than that for the reboot-based proactive recovery.

\begin{figure*}[t]
\begin{center} 
\leavevmode
\epsfxsize=6.0in
\epsfbox{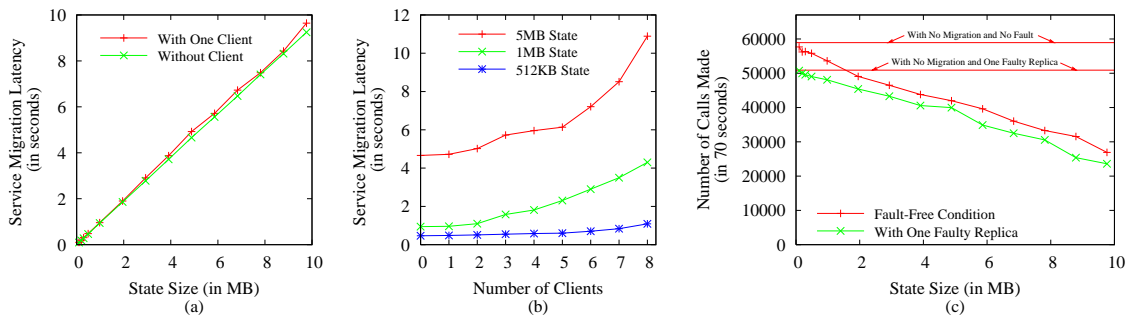}
\caption{(a) Service migration latency for different state sizes measured
when (1) the replicas are idle (other than the service migration activity),
labeled as ``Without Client'' and (2) in the presence of one client.
(b) The impact of system load on the migration latency.
(c) The impact of proactive migration on system performance (without and with
one crashed replica), as observed by a single client in terms of the number of 
calls made within one vulnerability window (70$s$).}
\label{perfig}
\end{center}
\end{figure*}

\section{Performance Evaluation}
\label{perfsec}
The proactive service migration mechanisms 
have been implemented and incorporated into the BFT framework developed by 
Castro, Rodriguos and Liskov~\cite{bft-osdi99, bft-osdi2000, bft-acm, base}. 
Due to the potential large state, an optimization has been made, similar to 
the optimization on the reply messages in the original BFT framework, 
\ie instead of
every replica sends its checkpoint to the target nodes of migration,
only one actually sends the full checkpoint. The target node can verify if the
copy of the full checkpoint is correct by comparing the digest of the 
checkpoint with the digests received from the replicas. If the checkpoint is 
not correct, the target node asks for a retransmission from other replicas.

Similar to \cite{bft-acm}, the performance measurements are carried out
in general-purpose servers without hardware coprocessors.
The related operations are simulated in software. Furthermore,
the trusted configuration manager is not developed as this is
one of the no goals of this paper. The motivation of the measurements
is to assess the runtime performance of the proactive service migration
scheme for practical use.
 
Our testbed consists of a set of Dell SC440 servers connected by a 
100 Mbps local-area network. Each server is equipped with a single
Pentium dual-core 2.8GHz CPU and 1GB of RAM, and runs the SuSE Linux
10.2 operation system. The micro-benchmarking example included in
the original BFT framework is adapted as the test application.
The request and reply message length is fixed
at 1KB, and each client generates requests consecutively in a loop
without any think time. Each server replica
simply echoes the payload in the request back to the client.

Four active nodes, four standby nodes, and up to eight client nodes 
are used in the experiment. This setup can tolerate a single
Byzantine faulty replica. The service migration interval is set to 70$s$,
corresponding to the minimum possible vulnerability window for a key
exchanged interval of 15s and a maximum recovery time (for a single replica)
of 10$s$.

To characterize the runtime cost of the service migration scheme, we measure 
the recovery time for a single replica with and without the presence of
clients, and the impact of proactive migration on the system performance
perceived by clients. The recovery time is determined by measuring the time 
elapsed between the following two events: (1) the primary sending the 
{\sc pre-prepare} message for the {\sc migration-request}, and
(2) the primary receiving a notification from the target standby node 
indicating that it has collected and applied the latest stable checkpoint. 
(The notification message is not part of the recovery protocol. It is
inserted solely for the purpose of performance measurements.) We refer
to this time interval as the service migration latency. The impact on the 
system performance is measured at the client by counting the number of
calls it has made during one vulnerability window, with and without proactive
migration-based recovery.

The measurement results are summarized in Figure~\ref{perfig}. 
Figure~\ref{perfig}(a) shows the service migration latency for various
state sizes (from 100KB to about 10MB). It is not surprising to see that the 
cost of migration is limited by the bandwidth available (100Mbps) because
in our experiment, the time it takes to take a local checkpoint (to
memory) and to restore one (from memory) is negligible. This is intentional
for two reasons: (1) the checkpointing and restoration cost is very
application dependent, and (2) such cost is the same regardless of the
proactive recovery schemes used. 

Furthermore, we measure the migration
latency as a function of the system load in terms of the number of 
concurrent clients. The results are shown in Figure~\ref{perfig}(b).
As can be seen, the migration latency increases more significantly for 
larger state when the system load is higher. When there are eight
concurrent clients, the migration latency for a state size of 5MB exceeds
10$s$, which is the maximum recovery time we assumed in our availability
analysis. This observation suggests the need for dynamic adjusting
of some parameters related to the vulnerability window, in particular,
the watchdog timeout used in the reboot-based recovery scheme. If the
watchdog timeout is too short for the system to go through four rounds
of proactive recovery (of $f$ replicas at a time), there will be more
than $f$ replicas going through proactive recoveries concurrently, which will 
decrease the system availability, even without any fault.
Our migration-based proactive recovery does not suffer from this problem.
Due to the use of coordinated recovery, when the system load increases, the
vulnerability window automatically increases.

Figure~\ref{perfig}(c) shows the performance impact of proactive service
migration as perceived by a single client. In the experiment, we choose
to use the parameters consistent with those used in the availability
analysis (for migration-based recovery), \ie key exchange period of 15$s$,
maximum recovery time of 10$s$, and a vulnerability window of 70$s$.
As can be seen, the impact of proactive migration on the system performance
is quite acceptable. For a state smaller than 1MB, the throughput is reduced
only by 10\% or less comparing with the no-proactive-recovery case. 
In addition, we have measured the migration performance in the presence of
one (crash) faulty replica (the recovering recovery is different from
the crashed replica). The system throughput degradation is similar to that
in fault-free condition. Note that when there are only three correct replicas,
the system throughput is reduced even without proactive migration, as shown
in the figure.

\section{Related Work}

Ensuring Byzantine fault tolerance for long-running systems
is an extremely challenging task. The pioneering work is carried
out by Castro and Liskov. In~\cite{bft-acm}, they proposed a
reboot-based proactive recovery scheme as a way to repair compromised
nodes periodically. The work is further extended by Rodrigues 
and Liskov in~\cite{bftlls}. They proposed additional infrastructure
support and related mechanisms to handle the cases when a damaged replica
cannot be repaired by a simple reboot. Our work is inspired by both work.
The novelty and the benefits of our service-migration scheme 
over the reboot-based proactive recovery scheme have been elaborated in 
Section~\ref{benefitsec}. 

Other related work includes~\cite{pallemulle}. In~\cite{pallemulle}, 
Pallemulle {\em et al.} extended the BFT algorithm to handle replicated 
clients and 
introduced another Byzantine agreement (BA) step to ensure that all replicated 
clients receive the same set of replies. It was claimed that the mechanisms 
can also be used to perform online upgrading, which is important for 
long-running applications and not addressed in our work. However, it is not
clear if the BA step on the replies is useful, while incurring significantly 
higher cost. If there are more than $f$ compromised server replicas, the
integrity of the service is already broken, in which case, there is no use
for the client replicas to receive the same faulty reply.

Finally, the reliance on extra nodes beyond the $3f+1$ active nodes in our
scheme may somewhat relates to the use of $2f$ 
additional witness replicas in the fast Byzantine consensus 
algorithm~\cite{fbc}. However, the extra nodes are needed for completely
different purposes. In our scheme, they are required for proactive recovery
for long-running Byzantine fault tolerant systems. In \cite{fbc}, however,
they are needed to reach Byzantine consensus in fewer message delays.

\section{Conclusion}
In this paper, we presented a novel proactive recovery scheme based
on service migration for long-running Byzantine fault tolerant
systems. We described in detail the challenges and mechanisms needed
for our migration-based proactive recovery to work. The migration-based 
recovery scheme has a number of unique benefits over previous work,
including a smaller vulnerability window by shifting the time-consuming 
repairing step out of the critical recovery path, higher system availability 
under faulty conditions, and self-adaptation to different system loads.
We validated these benefits both analytically and experimentally. For
future work, we plan to investigate the design and implementation of the 
trusted configuration manager, in particular, the incorporation of the 
code attestation methods \cite{code1,code2} into the fault detection 
mechanisms, and the application of the migration-based
recovery scheme to practical systems such as networked file systems.

\end{document}